\documentclass[preprint,12pt]{aastex}
\usepackage{epsfig, amsmath, amsfonts}

\newcommand\bb[1] {   \mbox{\boldmath{$#1$}}  }

\newcommand\del{\bb{\nabla}}
\newcommand\bcdot{\bb{\cdot}}
\newcommand\btimes{\bb{\times}}

\newcommand\kva{ \bb{k\cdot v_A}  }

\newcommand\zb{ {Z_b} }

\def\dd{\partial}

\def\beq{ \begin{equation} }
\def\eeq{ \end{equation} }
\def\spose#1{\hbox to 0pt{#1\hss}}  
\def\ltsim{\mathrel{\spose{\lower.5ex\hbox{$\mathchar"218$}}
\raise.4ex\hbox{$\mathchar"13C$}}}
\def\gtsim{\mathrel{\spose{\lower.5ex\hbox{$\mathchar"218$}}
\raise.4ex\hbox{$>$}}}

\slugcomment{Version: \today}

\begin{document}

\title{\bf\LARGE Dynamics of the Magnetoviscous Instability}
\author{Tanim Islam\altaffilmark{1} \& Steven Balbus\altaffilmark{1,2}}
\altaffiltext{1}{Department of Astronomy, P.O. Box 3818, University
  of Virginia, Charlottesville, VA 22903; \texttt{tsi6a@virginia.edu}}
\altaffiltext{2}{Laboratoire de Radioastronomie, \'Ecole Normale
Sup\'erieure, 24 rue Lhomond, 75231 Paris, France;
  \texttt{steve.balbus@lra.ens.fr}}

\begin{abstract}
In dilute astrophysical plasmas, the collisional mean free path of a
particle can exceed its Larmor radius.  Under these conditions, the
thermal conductivity and viscosity of the plasma can be dramatically
altered.  This alteration allows outwardly decreasing angular velocity
or temperature gradients to become strongly destabilizing.  This paper
generalizes an earlier, simple analysis of the viscosity instability,
by including the dynamical effects of magnetic field line tension.
Such effects lower the growth rates found in the absence of such tension,
but still allow growth rates in excess of the maximum of the standard
magnetorotational instability.  We find very good quantitative agreement
with more complex kinetic treatments of the same process.  The combination
of large growth rates and large magnetic Prandtl number suggest that
protogalactic disks are powerful dynamos.

\end{abstract}

\keywords{accretion, accretion disks; magnetic fields; MHD; instabilities;
galaxies: magnetic fields.}

\maketitle

\section{Introduction}
Magnetic fields, even when highly subthermal, turn free energy
gradients in sufficiently ionized fluids into sources of instability.
There are important astrophysical consequences of this result:
by the magnetorotational instability (MRI; Balbus \& Hawley 1991),
accretion disks become turbulent when angular velocity $\Omega$,
rather than angular momentum $\Omega R^2$, decreases outward, and by the
magnetothermal instability (MTI; Balbus 2001), dilute, stratified plasmas
are destablized when temperature rather than entropy decreases upwards.
Since it is easier to violate the free energy gradient criteria than
the classical Rayleigh and Schwarzschild criteria, it is the former,
when applicable, that are relevant to the behavior of their host systems.

The MRI is the best-known example of a free-energy gradient instability,
and has by now been extensively studied.  Recently, however, it has been
noted that purely viscous effects in a dilute MHD fluid can also lead
to accretion disk turbulence, even if the ${\bb{J\times B}}$ Lorentz
term ($\bb{J}$ is the current density and $\bb{B}$ is the
magnetic density)
is negligible (Balbus 2004; see Quataert et al. 2002 and Sharma
et al. 2003 for plasma kinetic treatments).  This is because there is
a large parameter domain in which the ion Larmor radius is very small
compared with fluid length scales (bounding the field strength from
below), yet the magnetic field contribution to fluid stresses is tiny
(bounding the field strength from above).  If the plasma in question
is sufficiently dilute, the ion cyclotron period is much smaller than
the ion-ion collision time.  Under these circumstances, the viscous
stress tensor is highly anisotropic, and dominated by its diagonal terms
\citep{b65}.  These circumstances are similar to those of anisotropic
thermal conduction.   In the MTI, heat, borne by the electrons, flows
only along the magnetic lines of force.  In the viscous case, it is of
course angular momentum flow that is restricted.  In both cases, fluid
attributes that are ordinarily responsible for stabilizing dissipation
(thermal conduction, viscosity) become active agents of destabilization,
a remarkable turn of events.

The cause of the instability is essentially the same in both cases.
Perturbed magnetic field lines, initially isothermal (isorotational),
are stretched along the direction of the background temperature (angular
velocity) gradient.  This allows heat (angular momentum) to flow from
one fluid element to another.  The elements then move yet farther apart,
the field lines become more aligned with the background gradient, and the
preocess runs away.  
The magnetothermal instability has been followed calculated far into the
nonlinear regime, in which vigorous convection has developed \citep{ps05}.

Neglect of the fluid $\bb{J\times B}$ force in the magnetoviscous
calculation of Balbus (2004), while enabling a point of principle to be
made more transparently, ultimately restricts the domain of validity.
Magnetic stresses are an important complication, acting as both a
stabilizing and destabilizing agent.  To understand the true nature
of this instability in an astrophysical context, it is desirable to
pursue a more general approach, and this is the primary motivation of
the current paper.

In \S 2, we discuss the physical parameter regime in which
this work is expected to apply, and present the formalism by which the
components of the magnetic viscosity may be calculated.
In \S 3, we formulate the problem and obtain the dispersion
relation of interest.  Characteristic growth rates are
determined, and the physical regime reexamined for self-consistency.
In \S 4, we present a discussion relating our work to
a more complex plasma kinetic treatment, and touch upon some
astrophysical implications.
Finally, in \S V, we summarize our conclusions.

\section {Preliminaries}
\subsection {Parameters}

The magnetoviscous instability finds its natural venue in protogalactic
disks and halos, as well as in very low density accretion flows, of
which the Galactic center is the prototype (Melia \& Falcke 2000).
These systems are characterized by
subthermal magnetic fields in collisionless plasmas and
ohmic diffusion coefficients small relative to the viscous
diffusion coefficient.  Let $\omega_{ci}$ be the ion
cyclotron frequency, and $\tau_{i}$ the ion-ion collision time:
\beq
\tau_i \simeq 5.91\times 10^5\,
\left(\frac{20}{n\ln\Lambda}\right)T_4^{3/2}
\ {\rm s} ,
\eeq
where $n$ is the proton
density in cm$^{-3}$, $T_4$ the ion kinetic temperature in units of $10^4$
K
and $\ln \Lambda$ is Coulomb logarithm.
We shall work in the
asymptotic domain
\begin{equation}
\omega_{ci} \tau_{i} = \left(\frac{1.09\times 10^5}{n}\right)
\frac{T_4^{3/2}B_{\mu G}}{\ln \Lambda} \gg 1 ,
\eeq
where $B_{\mu G}$ is the magnetic field strength in microgauss.
With $n \lesssim 1$ and $T_4 \gtrsim
1$, even very weak fields can be accommodated by this regime.
On the other hand, we shall also assume that the
Reynolds number is large, which requires
\beq
\Omega \tau_{i} \ll 1 ,
\eeq
where $\Omega$ is the disk rotational frequency.   The collision frequency
$1/\tau_i$ is therefore much larger than an orbital frequency, but much
smaller than the cyclotron frequency.  Finally, let us note the
Prandtl number ratio of the viscosity to ohmic resistivity
(Balbus \& Hawley 1998):
\beq
{\cal P} = \left(\frac{T}{10^4}\right)^4\left(\frac{6.5\times 10^{10}}{n}\right)
\left(\frac{20}{\ln \Lambda}\right)^2 .
\eeq
We shall assume ${\cal P} \gg1$.
The Spitzer (1962) viscosity for a hydrogenic plasma
is:
\beq\label{visc}
\eta \simeq  1.1 \times10^{-16} T^{5/2}\ \left(\frac{20}{\ln\Lambda}\right)
\mbox {g cm$^{-1}$ s$^{-1}$}.
\eeq
(Applications to low luminosity black hole accretion should use a value
of $\ln \Lambda$ closer to 30.)
Finally, for future reference, we use the standard plasma $\beta$
parameter to represent the ratio of the gas to magnetic pressures:
\beq
\beta = \frac{8\pi P}{B^2} .
\eeq

It is helpful to have representative physical parameters at hand, even if
they are very crude.  For protogalactic disks, densities could range from
$10^{-2}$ to 1 particle per cm$^3$; temperatures from $10^4$ to $10^6$ K.
For radiatively inefficient black hole accretion flows, a typical density
might be $10^8$ cm$^{-3}$, but the range of interest should be thought of
as perhaps $10^5-10^{10}$.   The ion and electron temperatures seem to be
very different in these flows: the ions ought to be virialized at
$T\sim 10^{12}$ K, whereas the electrons are thought to be ${}\ltsim 10^{10}$ K
(Narayan, Mahadevan, \& Quataert 1998 for a review).
In a forthcoming paper, we shall discuss a possible reason for this
two temperature structure based on the MTI.

\subsection{Magnetic Viscosity}

A convenient formalism for the viscous stress tensor in the
presence of a magnetic field is presented in Balbus (2004),
and shall quote the results here for reference, referring the
reader to this paper for further details.

Our fundamental coordinate system for the disk will be a
standard cylindrical system: radius $R$, azimuth $\phi$,
and axial variable $Z$.  Define now a local Cartesian system,
determined by the magnetic field.  We denote these axes by
subscript $b$.  $Z_b$ points along the local direction of the
magnetic field, $X_b$ and $Y_b$ may be any axes orthogonal
to one another as well as to the $Z_b$ direction.
Following Braginskii (1965), the $Z_b Z_b$ component of the
viscous stress $\sigma_{Z_b Z_b}$
is unaffected by the presence of the field.
It is given by (Balbus 2004):
\beq\label{sigzz}
\sigma_{Z_b Z_b} = - 2\eta\left[ (\bb{b\cdot\del})\bb{v}\right]
\bcdot \bb{b} , 
\eeq
where
$\bb{v}$ the local velocity field, and
$\bb{b}$ is a unit vector in the direction of the
magnetic field.  The other diagonal components of the
traceless viscous stress are
\beq\label{sigxxyy}
\sigma_{X_b X_b} = \sigma_{Y_b Y_b} = -{\sigma_{Z_b Z_b}/2} .
\eeq
To find the components of the magnetized viscous stress tensor
in any other
local Cartesian frame, the transformation law may be written
\beq
\sigma_{ij} = \sum_{i_b, j_b}
(\bb{i}\bcdot \bb{i_b})\ (\bb{j}\bcdot\bb{j_b})\ \sigma_{{i_b}
{j_b}},
\eeq
where once again the $b$ subscript denotes the magnetic field
frame and
bold face quantities are unit vectors of the indicated
component. Using
equation (\ref{sigxxyy}) for the nonvanishing diagonal stress
tensor
components in the field frame, we obtain
\beq\label{sigij}
\sigma_{ij} = \sigma_{\zb\zb}\left[
( \bb{i}\bcdot \bb{Z_b})(\bb{j}\bcdot\bb{Z_b})
- \frac{1}{2}( \bb{i}\bcdot \bb{Y_b})(\bb{j}\bcdot\bb{Y_b})
-\frac{1}{2}( \bb{i}\bcdot \bb{X_b})(\bb{j}\bcdot\bb{X_b})
\right].
\eeq
Once $\sigma_{Z_b Z_b}$ is determined in the magnetic field
frame, $\sigma_{ij}$ may be calculated in any frame.

\section{Formulation of the Problem}
\subsection{Braginskii Stress}

We consider the stability of a disk under the influence of a
weak magnetic
field.  As in standard MRI analyses, we assume that the field
is sufficiently weak that it has no effect on the equilibrium
state, but that magnetic tension is important for the behavior of
local WKB perturbations.  The fundamental fluid equations used here are
mass conservation,
\beq
\frac{\dd \rho}{\dd t} + \del\bcdot (\rho \bb{v})= 0,
\eeq
the equation of motion,
\beq
\rho\left(\frac{\dd}{\dd t} + \bb{v}\bcdot\del \right)\bb{v} =
-\del \left( P + \frac{B^2}{8\pi}\right)
+\frac{1}{4\pi} \bb{(B\cdot\nabla)B} - \rho\del\Phi
-\frac{\dd\sigma_{ij}}{\dd x_j},
\eeq
and the induction equation of ideal MHD,
\beq
\frac{\dd\bb{B}}{\dd t} = \del\btimes(\bb{v}\btimes \bb{B}).
\eeq
$\Phi$ represents an external gravitational potential
and the other symbols have their usual meanings; the viscous
stress tensor $\sigma_{ij}$ is given by equation (\ref{sigij}).

The equilibrium state is a differentially rotating disk.  As
stated above, we work
in a standard cylindrical coordinate system, $R, \phi, Z$.  The angular
velocity is $\Omega(R)$, and we shall restrict ourselves to a local
analysis at the midplane.  Thus, we may ignore buoyant forces.

In the equilibrium state, it is assumed that $\sigma_{ij} =0$;
it will be shown that this is in general not a stable configuration.
The initial magnetic field lines are spooled around cylinders,
unaffected by the shear.

We consider next small departures from the equilibrium flow.  Linearly
perturbed quantities are denoted by $\bb{\delta v}, \delta\sigma_{ij}$,
etc.  We work in the local WKB limit, with the space-time dependence of
all perturbed quantities given by $\exp(\gamma t + i\bb{k}\bcdot\bb{r})$.
Thus, $\gamma$ is a growth or decay rate if it is real, and an angular
frequency if it is imaginary.

To evaluate $\delta\sigma_{ij}$ we follow Balbus (2004).
Since $\sigma_{ij}$
vanishes in the equilibrium state, it follows from equation (\ref{sigij})
that
\beq
\delta \sigma_{ij} = \delta\sigma_{\zb\zb}\left[
( \bb{i}\bcdot \bb{Z_b})(\bb{j}\bcdot\bb{Z_b})
- \frac{1}{2}( \bb{i}\bcdot \bb{Y_b})(\bb{j}\bcdot\bb{Y_b})
-\frac{1}{2}( \bb{i}\bcdot \bb{X_b})(\bb{j}\bcdot\bb{X_b})
\right].
\eeq
The geometry of the equilbrium field is defined by
\beq
\bb{b} = \cos\chi\  \bb{{\hat \phi}} + \sin\chi\ \bb{{\hat Z}}.
\eeq
$\bb{{\hat \phi}}$, $\bb{{\hat Z}}$, and $\bb{{\hat R}}$
(used below) are unit vectors in the indicated cylindrical
directions, and $\chi$ is the angle between the magnetic
field and the $\phi$ axis.  The local magnetic field axes are
chosen to be
\beq\label{b}
\bb{Z_b} = \bb{b} = \cos\chi\ \bb{{\hat \phi}} + \sin\chi\ \bb{{\hat Z}},
\eeq
\beq
\bb{X_b} = \bb{{\hat R}}\btimes \bb{Z_b} =
-\sin\chi\  \bb{{\hat \phi}} + \cos\chi\ \bb{{\hat Z}},
\eeq
and
\beq
\bb{Y_b} = \bb{{\hat R}}.
\eeq
See fig.\ 2.

The diagonal elements of $\delta\sigma_{ij}$ are found to be
\beq
\delta\sigma_{RR} = -\frac{1}{2}\delta\sigma_{\zb\zb}, \
\delta\sigma_{\phi\phi} = \left (\cos^2\chi - \frac{\sin^2\chi}{2}
\right) \delta\sigma_{\zb\zb}, \
\delta\sigma_{ZZ} = \left(\sin^2\chi -\frac{\cos^2\chi}{2}\right)
\delta\sigma_{\zb\zb},
\eeq
and the off-diagonal elements are
\beq
\delta\sigma_{\phi Z}=
\delta\sigma_{Z\phi}= \frac{3}{2}\cos\chi\ \sin\chi\ \delta\sigma_{\zb\zb} .
\eeq
Finally, we evaulate $\delta\sigma_{\zb\zb}$ from its vector-invariant form
(\ref{sigzz}),
\beq
\delta\sigma_{\zb\zb} = -2 \eta
\left(
\left[(\bb{\delta b}\bcdot\del)\bb{v}\right]\bcdot\bb{b}+
\left[(\bb{b}\bcdot\del)\bb{\delta v}\right]\bcdot\bb{b}+
\left[(\bb{b}\bcdot\del)\bb{v}\right]\bcdot\bb{\delta b}
\right) .
\eeq
Using $\bb{v}=R\Omega\bb{\hat \phi}$ and equation (\ref{b}), we obtain
\beq\label{sigzz2}
\delta\sigma_{\zb\zb} =-2\eta \left[ \delta b_R\, \frac{d\Omega}{d\ln R}\, \cos\chi +
i(\bb{k}\bcdot\bb{b})(\bb{b}\bcdot\bb{\delta v})\right].
\eeq

\subsection{Linearized Equations}
The linearized dynamical equations are
\beq
\bb{k}\bcdot\bb{\delta v} = 0,
\eeq
\beq
\gamma \delta v_R -2\Omega \delta v_\phi +
\frac{ik_R}{\rho}\left( \delta P -  \frac{\delta\sigma_{\zb\zb}}{2}
+\frac{\bb{B\cdot\delta B}}{4\pi}
\right) - \frac{ik_Z B_Z\sin\chi}{4\pi\rho} \delta B_R = 0,
\eeq
\beq
\gamma \delta v_\phi + \frac{\kappa^2}{2\Omega} \delta v_R
 +i\frac{3k_Z}{2\rho} \sin\chi\, \cos\chi\, \delta
 \sigma_{\zb\zb}
 - \frac{ik_Z B_Z\sin\chi}{4\pi\rho } \delta B_\phi= 0,
\eeq
(Here, $\kappa^2$ is the square of the epicyclic frequency,
$\kappa^2 = 4\Omega^2 + {d\Omega^2 /d\ln R}$.)
\beq
\gamma\delta v_Z
+\frac{ik_Z}{\rho} \left[
\delta P +\left({\sin^2\theta } - \frac{\cos^2\theta}{2}\right)
\delta\sigma_{\zb\zb} \right]
 - \frac{ik_Z B_Z\sin\chi}{4\pi\rho } \delta B_Z= 0.
\eeq
The induction equations are
\beq
\gamma \delta B_R - i k_Z B_Z \sin\chi \delta v_R = 0,
\eeq
\beq
\gamma\delta B_\phi -\delta B_R \frac{d\Omega}{d\ln R} - k_ZB_Z\sin\chi
\delta v_\phi = 0,
\eeq
\beq
\gamma \delta B_Z - i k_Z B_Z \sin\chi \delta v_Z = 0.
\eeq
Finally, we recast equation (\ref{sigzz2}) for the all-important $Z_b Z_b$
stress component in a more convenient form,
\beq
\delta\sigma_{\zb\zb} = -2\eta i k_Z\sin\chi\left[
\cos\chi \left(
\frac{\delta v_R}{\gamma}\frac{d\Omega}{d\ln R} +\delta v_\phi\right)
+  \sin\chi\, \delta v_Z\right].
\eeq

\subsection {Dispersion Relation}

The linearized equations lead, after straightforward but
somewhat tedious algebra, to a dispersion relation that blends the
properties of the magnetorotational and magentoviscous instabilities:
\begin{eqnarray*}
\frac{k^2}{k_Z^2}\gamma^4 + 3\eta_V\gamma^3\sin^2\chi k_\perp^2
\left(k_R^2 + k_Z\cos^2\chi\right)
+\gamma^2\left(\kappa^2 + 2 \frac{k^2}{k_Z^2}(\kva)^2\right) +
\qquad\qquad\qquad\qquad\qquad\qquad\qquad &&
\end{eqnarray*}
\beq\label{main}
+3\eta_V\gamma\sin^2\chi\left[ (\kva)^2k_\perp^2 + k_Z^2\cos^2\chi
\frac{d\Omega^2}{d\ln R} \right]
+(\kva)^2 \left[ \frac{k^2}{k_Z^2}(\kva)^2 +
\frac{d\Omega^2}{d\ln R} \right] = 0,  {\qquad}
\eeq
where $k_\perp^2 = k_R^2 +k_Z^2\cos^2\chi$ is the square of the
component of
the wavenumber perpendicular to $\bb{b}$, and $\eta_V$ is the kinematic
viscosity $\eta/\rho$.  We note that in the limit $\eta_V\rightarrow0$
the standard MRI dispersion relation is recovered, while in the
limit $\kva\rightarrow0$, the magnetoviscous dispersion relation
of Balbus (2004) emerges (taking also the limit $k_R=0$).
Unstable modes are present if $d\Omega^2/dR<0$, as expected.

To evaluate the growth rates associated with the dynamical magnetoviscous
instability, we proceed as follows.  First, we set $k_R=0$,
which corresponds to the most rapidly growing modes, and simplifies the
calculation.  Second, we calculate all rates
($\gamma$, $\kva$, $\kappa$, $\eta_V k_\perp^2$)
in units of $\Omega$.  We introduce the notation
\beq
X\equiv (\kva)^2/\Omega^2, \qquad Y = 3\eta_V \sin^2\chi k_\perp^2/\Omega, 
\qquad \gamma'=\gamma/\Omega,
\eeq
and restrict ourselves to the astrophysically interesting
case of $d\Omega^2/dR<0$.  Our dispersion relation may then be rewritten as
\beq
Y = \frac{\gamma'^4 + \gamma'^2[(\kappa/\Omega)^2 + 2X] +X(X+d\ln\Omega^2/d\ln R)}
{\gamma' (|d\ln\Omega^2/d\ln R| - X-\gamma'^2) }.
\eeq
Triplets of the form $(X,Y,\gamma')$ are easily calculated by this formula,
and contour plots of $\gamma$ thereby generated.  In fact,
it is convenient to view the results in the $Y/X$, $X$ plane
since the wavenumber is thereby removed from the ordinate,
which becomes a normalized viscosity.
The results are shown in figures 1 (Keplerian rotation profile)
and 2 (galactic rotation profile).  The maximum possible growth rate is
\beq\label{gammax}
\gamma^2_{max} = \left|\frac{d\Omega^2}{d\ln R}\right| .
\eeq
as found by plasma kinetic treatments
(Quataert et al. 2002; Sharma et al. 2003) and for the magnetoviscous
instability (Balbus 2004).  The effect of magnetic tension is to
lower the maximum growth rate to somewhere between this value,
and the MRI maximum rate of $0.5|d\Omega/d\ln R|$.
\begin{figure}[!ht]\centering
    \epsfig{file = 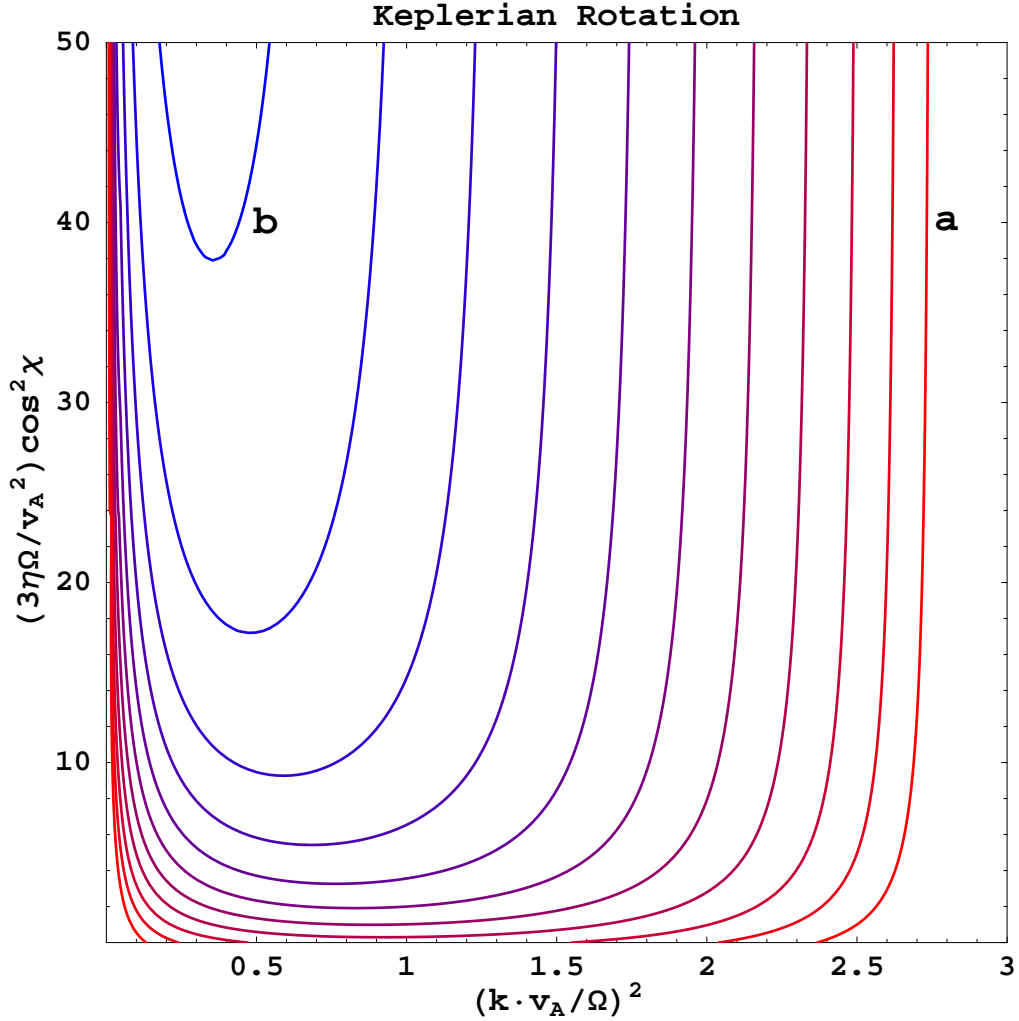, width = 0.82\linewidth}
    \caption{Contours of normalized growth rate $\gamma/\Omega$ for a
	Keplerian rotation profile, as a function of normalized wavenumber
	$\left({\bf k}\cdot{\bf v}_A/\Omega\right)^2$ and normalized viscosity
	$3\eta\Omega\cos^2\chi/v_A^2$. Contours run from $\gamma =
	0.5\Omega$ (contour a) to $\gamma = 1.5\Omega$ (contour b) in
	steps of $0.1\Omega$.  The maximum MRI growth rate is $0.75\Omega$.}
\end{figure}
\begin{figure}[!ht]
	\epsfig{file = 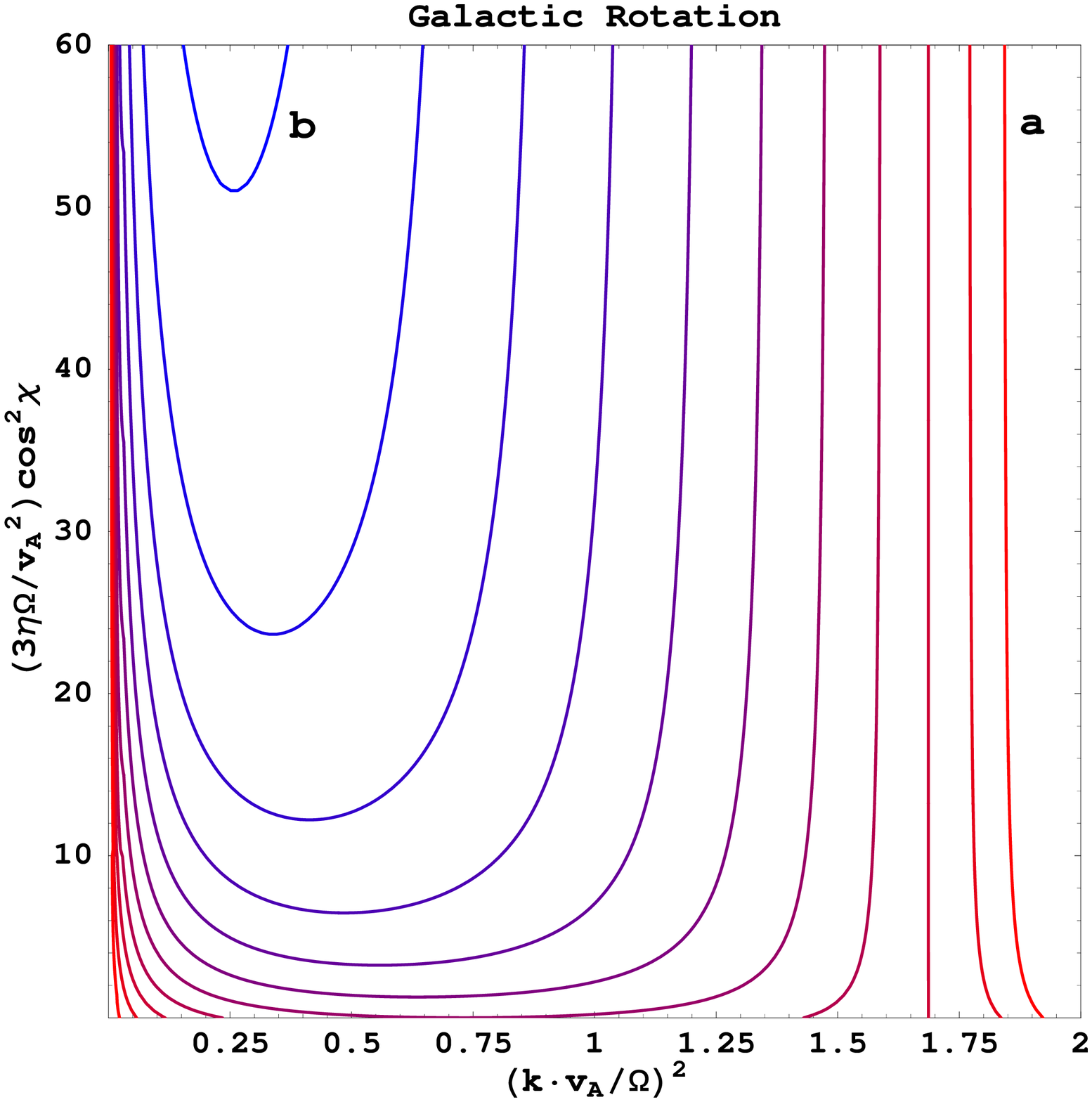, width = \linewidth}
	\caption{Contours of normalized growth rate $\gamma/\Omega$ for a
	galactic rotation profile ($\Omega\propto R^{-1}$), as a function
	of normalized wavenumber $\left({\bf k}\cdot{\bf v}_A/\Omega\right)^2$
	and normalized viscosity $3\eta\Omega\cos^2\chi/v_A^2$. Contours
	run from $\gamma = 0.4\Omega$ (contour a) to  $\gamma = 1.2\Omega$
	(contour b) in steps of $0.08\Omega$.  The maximum MRI growth
	rate is $0.5\Omega$.}
\end{figure}

\subsection {Physical Regime}

Figures (1) and (2) show that magnetoviscous effects can significantly
change the inviscid MRI growth rates in the parameter
regime $X\ltsim 1$, $Y\gtsim 1$.  Let us determine what sort of
magnetic field strengths are implied by this condition
in an environment representative of a protogalaxy.
With $\sin^2\chi$ set equal
to a half,
\beq
Y = \frac{1.5 \eta_V k_\perp^2}{\Omega} = \frac{1.5\eta_V\Omega}{c_S^2}
(k_\perp H)^2,
\eeq
where we have introduced the isothermal sound speed $c_S$ and
scale height $H$ defined by $H\Omega=c_S$.  Replacing $\eta_V$
by $\rho\eta$ and using equation (\ref{visc}), we find
\beq
Y = \frac{2.39 \times 10^{-8} T_4^{3/2}\Omega_{-14}}{n}\, (k_\perp H)^2 ,
\eeq
where $T_4$ is the temperature in units of $10^4$K, and $\Omega$
the angular rotation rate in units of $10^{-14}$ rad s$^{-1}$.
In other words,
\beq
(k_\perp H)^2 \gtsim 4.2\times 10^7 \frac{nT_4^{-3/2}}{\Omega_{-14}} ,
\eeq
for viscous effects to be important.  On the other hand, $X\ltsim 1$
for the growth rates to be significantly enhanced.  Since
\beq
X = \frac{k_Z^2 v_A^2}{\Omega^2} = (k_Z H)^2 \frac{v_A^2}{c_S^2} =
2 \frac{(k_Z H)^2}{\beta} ,
\eeq
we find that
\beq
\beta \gtsim 2 k_Z^2 H^2 \gtsim 8.4\times 10^7
\left(\frac{nT_4^{-3/2}}{\Omega_{-14}}\right)
\eeq
(ignoring the distinction between $k_Z$ and $k_\perp$.)  This translates
to a magnetic field strength of
\beq
B \ltsim 6.5\times 10^{-10} \ T_4^{5/4} \Omega_{-14}^{1/2} \quad
{\rm G} .
\eeq
This is of the order of the strength of the seed field
estimated in Balbus (2004), obtained by diluting
stellar surface mangetic fields of strength $\sim 0.1$ G over parsec scales.
For applications to inefficiently radiating black hole accretion flows,
the permissible $\beta$ parameter regime can extend to values near
unity.

\section {Comparison with Plasma Kinetic Treatment}

Anisotropic stresses in the presence of a magnetic field may be
calculated from a moment expansion of the collisionless Boltzmann
equation using a formalism developed by Kulsrud (1983).  (In the 
absence of a heat flux, this reduces to a double-adiabatic
equation of state first derived by Chew et al.\ [1956].)
This approach has been applied to
the MRI in the fully collisionless regime (Quataert et al. 2002) and
the collisional regime  (Sharma et al. 2003; see also Snyder, Hammett \& Dorland 1997).
In the fully collisional
limit, Sharma et al. recover the MRI, as do we in the limit
$\eta_V\rightarrow 0$.  The fluid and plasma kinetic treatments are
markedly different in their level of complexity, and a quantitative
comparison of their findings is therefore of interest.  This is shown
in figures 3 and 4.
\begin{figure}[!ht]
    \epsfig{file = 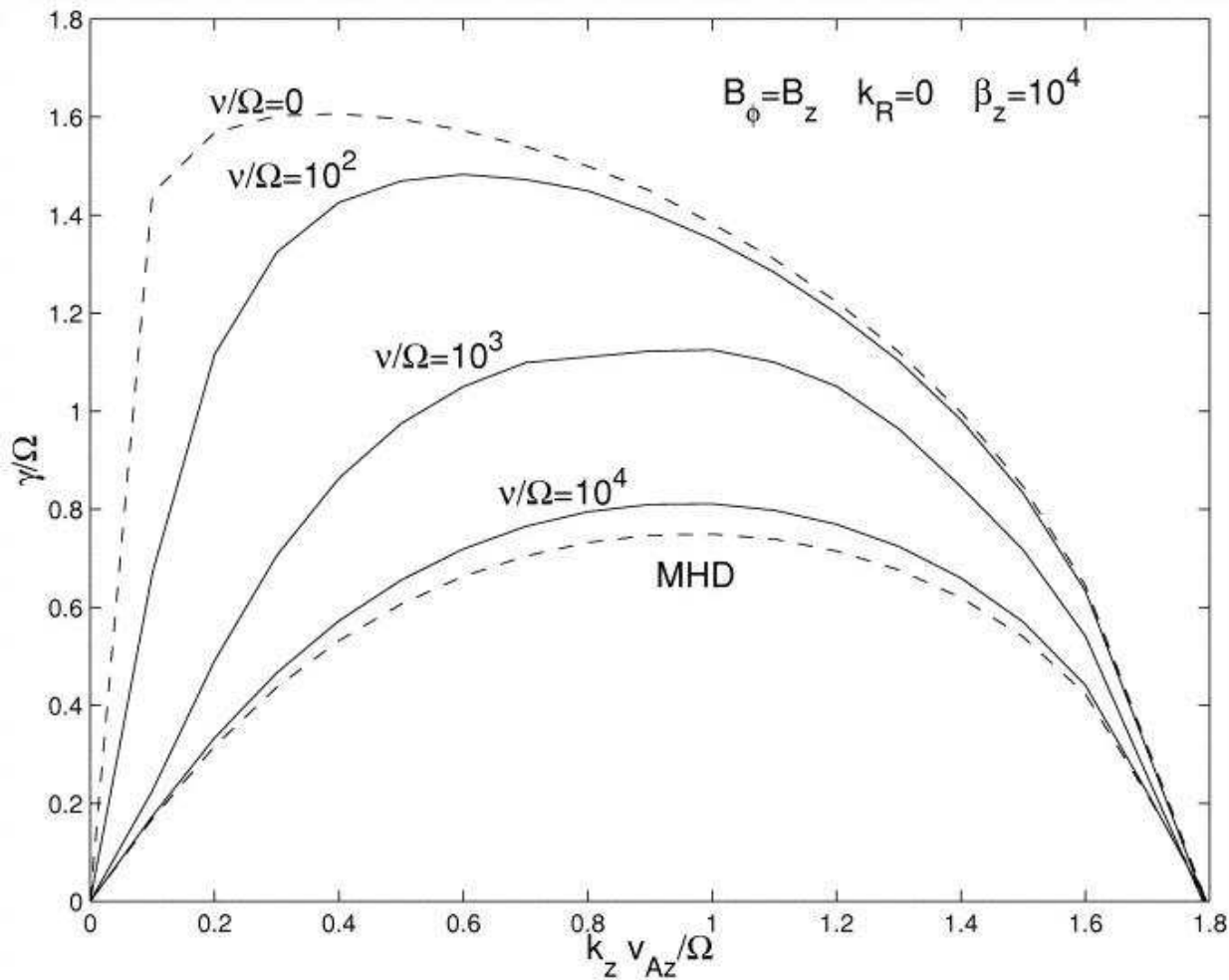, width = \linewidth}
    \caption{Dispersion relation for the magnetoviscous instability
    based on a kinetic treatment (figure adapted from Sharma et al. 2003).}
\end{figure}
\begin{figure}[!ht]
    \epsfig{file = 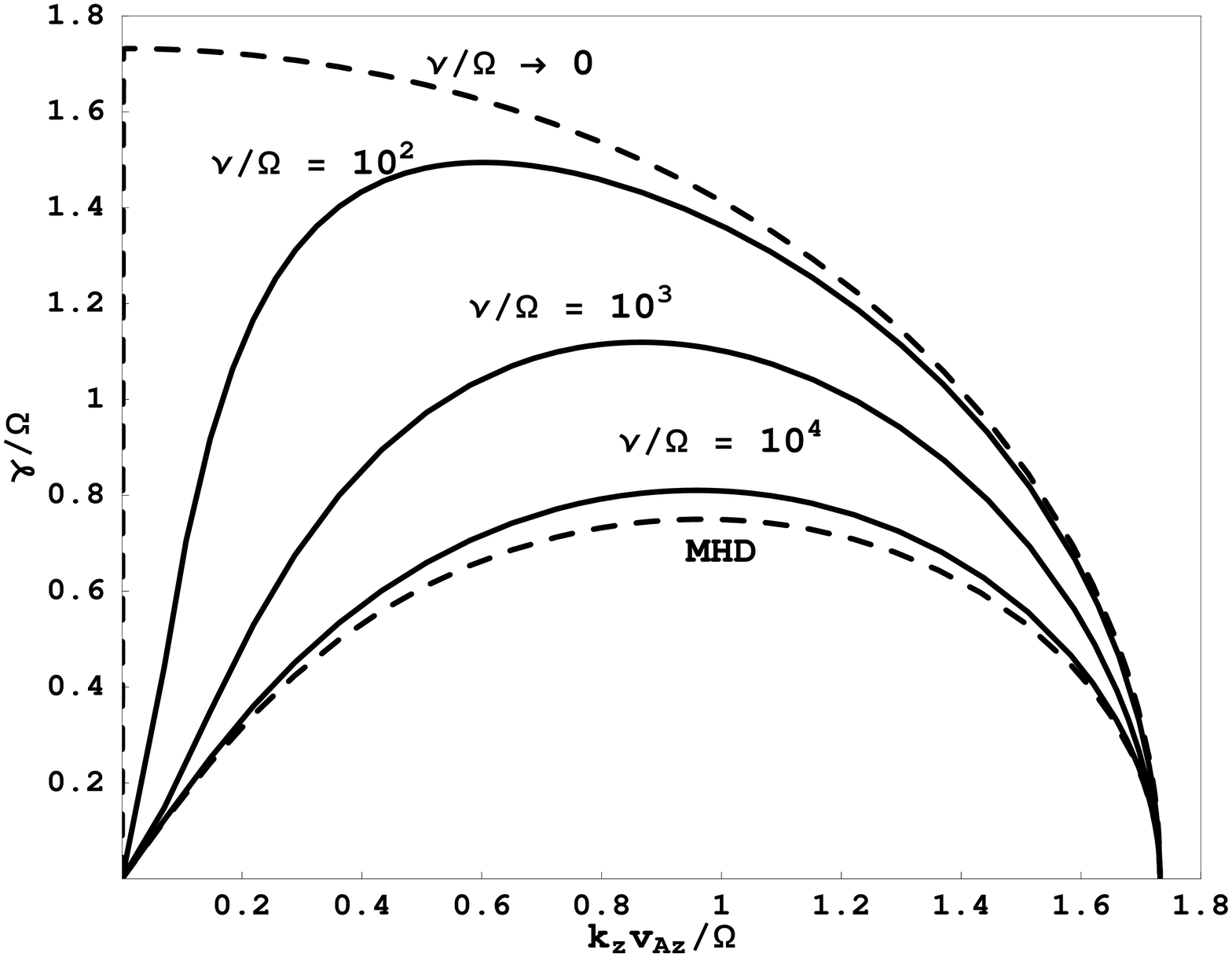, width = \linewidth}
    \caption{Dispersion relation for the magnetoviscous
    instability using an MHD fluid approach.}
\end{figure}
We have adopted the fiducial case of Sharma et al. (2003).
This corresponds
to $\chi = 45^\circ$,
$\beta_Z=8\pi P/B_Z^2 = 10^4$.  Choosing values for
$\kva/\Omega$ and $\nu/\Omega$, where $\nu$
is the ion self-collision frequency, then allows our
$X$ and $Y$ parameters to be determined,
once use is made of the Spitzer viscosity (\ref{visc}).
The figures show the growth rates as a function of $\kva/\Omega$
for several different values of $\nu/\Omega$.
In general the two approaches are in excellent agreement (growth
rates agree to within $\sim 5$), with
significant disparity coming only in the completely collisionless regime
(infinite viscosity).  This is not unexpected, since it corresponds to
the complete breakdown of the fluid regime.  What is remarkable is how
well the relatively simple fluid treatment performs.  The savings of
effort is the greater the more complex the problem.  For example, an
investigation combining both viscous and thermal effects is relatively
straightforward via a fluid treatment; both effects need to be included
in a study of high temperature accretion flows.  This will be presented
in a forthcoming paper by the authors.

It should be noted, however, that the effort entailed in a kinetic
treatment has valuable dividends.  
It is difficult to know, for example, 
how far into the long mean free path regime an MHD fluid approach
may be relied upon before
collective instabilities might be triggered.
This level of analysis can be provided only by a kinetic treatment
(e.g., Schekochihin et al. 2005).

\section {Conclusions}

We have analyzed the stability of a dilute astrophysical plasma in
the presence of Braginskii (1965) viscosity.  Our main result is the
dispersion relation (\ref{main}).  The regime of interest corresponds to
gases in which the ion gyroradius is small compared to a mean free path.
In laboratories this means a very strong field, but in astrophysical
applications even very weak fields are in this regime when the density is
sufficiently small.  This regime does not apply to classical accretion
disks or to stars (the densities are far too high), but it does apply
to the interstellar medium of galaxies and protogalaxies, and to the
low density accretion flows around some giant black holes (e.g. the
Galactic center).

The work presented here extends the study of Balbus (2004), which included
Braginskii viscosity but ignored $\bb{J\times B}$ Lorentz forces,
and complements the plasma kinetic treatment of Sharma et al. (2003).
The dynamical effects of the magnetic field lower the maximum growth rate
found in the absence of such effects (eq. [\ref{gammax}]), but growth
rates significantly in excess of the MRI value $0.5|d\Omega/d\ln R|$
are still found.

The magnetoviscous instability thus enhances the dynamo effect of the MRI
in dilute gases, and therefore may well be a significant source of
magnetic field amplifcation.  Whether it is a more efficient mechanism
than others that have been suggested (e.g.  Schekochihin et al. 2004)
is yet to be established, but numerical simulation of this large Prandtl
number regime is clearly now a viable possibility.  At the very least,
MRI growth rates are significantly enhanced in the linear regime, and
nonlinear field dissipation is inhibited (since the resistive scale is
``blocked'' in a downward cascade by dissipation at the larger viscous
scale).

In the presence of strong temperature gradients in 
accretion flows, magnetothermal and
magnetoviscous effects may both be important.  Since a mere temperature
gradient can now trigger convective instability, consequences for the
temperature structure of inefficiently radiating accretion flows can in
principle be profound.  These will be discussed in a forthcoming paper.

\section*{Acknowledgements}
It is a pleasure to acknowledge useful discussions with S. Cowley,
W. Dorland, and E. Quataert.   We thank the anonymous referee
for constructive comments leading to a greater clarity of presentation.   
This work was supported in part by the National Science Foundation under Grant No. 
PHY99--07949, and
by NASA Grants NAG5-13288 and NNG04GK77G.

\end{document}